\RequirePackage{fix-cm}
\documentclass[smallextended]{svjour3}       
\smartqed  
\usepackage{graphicx}
\begin{document}

\title{Characterizing the Interaction Between DNA and GelRed Fluorescent Stain.
}


\author{F. A. P. Crisafuli         \and
        E. B. Ramos                \and
        M. S. Rocha    
}


\institute{F. A. P. Crisafuli \at
              Laborat\'orio de F\'isica Biol\'ogica, Universidade Federal de Vi\c{c}osa, Minas Gerais,
              Brazil.
           \and
           E. B. Ramos \at
              Laborat\'orio de F\'isica Biol\'ogica, Universidade Federal de Vi\c{c}osa, Minas Gerais,
              Brazil.
           \and
           M. S. Rocha \at
              Laborat\'orio de F\'isica Biol\'ogica, Universidade Federal de Vi\c{c}osa, Minas Gerais,
              Brazil. \\
               Tel.: +55-31-38993399\\
              Fax: +55-31-38992483\\
              \email{marcios.rocha@ufv.br}
}

\date{Received: date / Accepted: date}

\maketitle

\begin{abstract}
We have performed single molecule stretching experiments and dynamic
light scattering (DLS) in order to characterize the interaction
between the DNA molecule and the fluorescent stain GelRed. The
results from single molecule stretching show that the persistence
length of the DNA-GelRed complexes increases as the ligand
concentration increases up to some critical concentration, then
decreasing for higher concentrations. The contour length of the
complexes, on the other hand, increases monotonically as a function
of GelRed concentration, suggesting that intercalation is the main
binding mechanism. In order to characterize the physical chemistry
of the interaction, we use the McGhee-von Hippel binding isotherm to
extract the physicochemical parameters of the interaction from the
contour length data. Such analysis has allowed us to conclude that
the GelRed stain is in fact a bis-intercalator. In addition, DLS
experiments were performed to study the changes of the effective
size of the DNA-GelRed complexes, measured by the hydrodynamic
radius, as a function of ligand concentration. We found a
qualitative agreement between the results obtained from the two
techniques by comparing the behaviors of the hydrodynamics radius
and the radius of gyration, since this last quantity can be
expressed as a function of mechanical parameters determined from the
stretching experiments.

\keywords{intercalation \and single molecule stretching \and dynamic
light scattering \and binding isotherm}
 \PACS{82.37.Rs \and 87.14.gk \and 87.80.Cc}
\end{abstract}

\section{Introduction}

Rational drug design is a process in which a new compound is
sketched and synthesized to achieve a specific biological target
and/or to perform specific functions. In health sciences, for
example, there is much interest in developing new drugs to treat
human diseases such as cancer. In biochemistry and molecular
biology, new drugs are usually developed to stain DNA and proteins,
allowing visualization by fluorescence microscopy and the
possibility of following the route of various biological processes.

GelRed is a fluorescent nucleic acid stain designed with the purpose
of replacing the highly toxic ethidium bromide (EtBr) in gel
electrophoresis and other experimental techniques which depends on
the fluorescence of stained DNA. When bound to DNA, GelRed has the
same absorption and emission spectra of EtBr and, according to its
manufacturer (Biotium Inc., Hayward, CA, USA), the compound can be
used in electrophoresis experiments with greater sensitivity than
EtBr, with the advantage of being  much less toxic and mutagenic
\cite{GRSafety}, \cite{Huang}. This last property, according to the
manufacturer, is due to the fact that the chemical structure of the
dye was designed in a way such that it is incapable of crossing cell
membranes \cite{GRSafety}. The chemical structure of the GelRed dye
is proprietary and was not officially reported by the manufacturer.
Nevertheless, one can find unofficial information in the internet
claiming that GelRed is synthesized basically by crosslinking two
EtBr molecules \cite{WikiGelRed}, which suggests that the dye is
maybe a bis-intercalator.

Even though the manufacturer states that GelRed binds to DNA via a
combination of intercalation and electrostatic binding
\cite{GRSafety}, most details of the interaction are so far not
reported in the literature. In this work we have performed single
molecule stretching experiments and dynamic light scattering in
order to gain insights about such interaction. Recently, we have
developed a methodology that allows one to easily extract the
physical chemistry of the DNA-ligand interaction from pure
mechanical DNA parameters such as those easily obtained from single
molecule stretching. This methodology allows one to determine the
relevant physicochemical parameters of the interaction and to deduce
the particular binding mechanism(s) \cite{Siman}, \cite{Cesconetto},
\cite{Silva}. The purpose of the present work is to perform a robust
characterization of the DNA-GelRed interaction, determining the
changes of the basic mechanical properties of the DNA molecule as
GelRed binds, the physicochemical parameters of the interaction and
the nature of the binding mechanism.

\section{Materials and Methods}

\subsection{Stretching experiments}

In these experiments the samples consist of $\lambda$-DNA molecules
end-labeled with biotin in a phosphate-buffered saline (PBS)
solution with [NaCl] = 140 mM. One end of the DNA molecule is
attached to a streptavidin-coated glass coverslip using the
procedure reported in ref. \cite{Amitani} while the other end of the
molecule is attached to a streptavidin-coated polystyrene bead with
3 $\mu$m diameter (Bangs Labs). As described earlier
\cite{CrisafuliIB}, \cite{Cesconetto}, in this configuration one can
easily trap the polystyrene bead with the optical tweezers and
stretch the DNA molecule by moving the microscope stage with a
piezoelectric actuator. The sample chamber consists of an o-ring
glued in the coverslip, such that one can exchange the buffer and
consequently the ligand concentration without affecting the trapped
DNA molecule by using micropipettes. The DNA base-pair concentration
used in all stretching experiments was C$_{bp}$ = 2.4 $\mu$M.

The optical tweezers consist of a 1064 nm ytterbium-doped fiber
laser with a maximum output power of 5.8 W (IPG Photonics) mounted
on a Nikon Ti-S inverted microscope with a 100 $\times$ N.A. 1.4
objective. The apparatus is previously calibrated by two independent
methods as described earlier \cite{CrisafuliIB}. Once calibrated,
the optical tweezers are used to trap the polystyrene bead attached
to a DNA molecule, allowing one to perform the DNA stretching with
high resolution and consequently to obtain the force $\times$
extension curves of the DNA-ligand complexes for each studied
situation. These curves were fitted to the Marko-Siggia WormLike
Chain (WLC) model \cite{Marko}. Therefore, one can study how the
persistence and contour lengths of the DNA molecule vary as a
function of ligand concentration. All the stretching experiments
were performed in the entropic low-force regime ($\leq$ 2 pN), in
order to avoid enthalpic contributions to the values of the
mechanical parameters \cite{CrisafuliIB}, \cite{Cesconetto}. All the
experimental details can be found in our references
\cite{CrisafuliIB}, \cite{Cesconetto}, \cite{Silva}, \cite{Reis}.

\subsection{Dynamic Light Scattering (DLS)}

In order to confirm the results obtained from the stretching
experiments with a second technique, we have also performed DLS on
the DNA-GelRed complexes. DLS performed at a fixed scattering angle
allows one to evaluate the hydrodynamic radius of the complexes, a
parameter that gives an estimation of the size of the complexes
\cite{Reis}. This parameter was measured in order to investigate the
effect of the ligand on the effective size of the DNA molecule,
which depends on its persistence and contour lengths. Thus, the
results obtained from two very different experimental techniques can
be compared (at least indirectly).

The DLS apparatus used was a ZetaSizer Nano-S (Malvern Instruments
Ltd) with a low volume cuvette (ZEN2112, Hellma Analytics). The
backscattering angle used was 173$^o$ in all experiments. The
particle size was determined using the Non-Negative Least Squares
(NNLS) algorithm.

The DNA used in these experiments was a 3000 bp molecule (Thermo
Scientific) in the same buffer used in the optical tweezers samples
($\lambda$-DNA is difficult to be used in DLS due to the long
contour length). The DNA molecules are equilibrated with a certain
concentration of GelRed directly in the cuvette used in the DLS
apparatus. The DNA concentration used in all DLS experiments was the
same one used in the optical tweezers (2.4 $\mu$M of base-pairs).
This concentration is sufficiently low to avoid entanglements and
relevant interactions between different DNA molecules \cite{Hur}.

\section{Results}

\subsection{Stretching experiments}

\begin{figure}
\includegraphics[width=8.5cm]{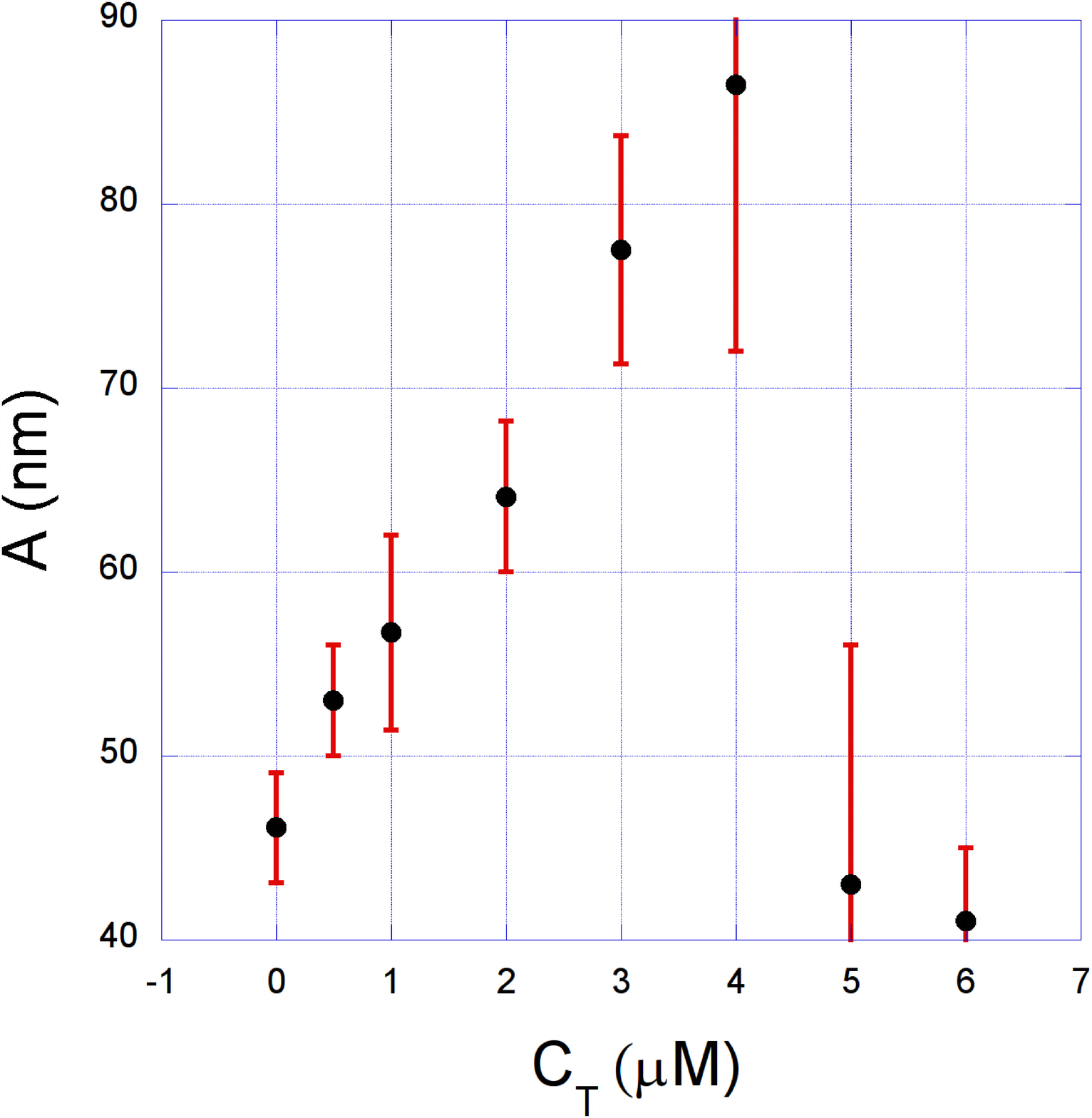}
\caption{Persistence length $A$ as a function of ligand total
concentration in the sample $C_T$. Observe that it initially
increases from the bare DNA value ($\sim 46$ nm) until reaching a
maximum value of $\sim 86$ nm at C$_{T}$ = 4.0$\mu$M. For higher
concentrations, it abruptly decays to $\sim 43$ nm.}\label{AxCt}
\end{figure}

In Fig. \ref{AxCt} we show the behavior of the persistence length
$A$ of DNA-GelRed complexes as a function of ligand total
concentration in the sample $C_T$. Observe that it initially
increases from the bare DNA value ($\sim 46$ nm) until reaching a
maximum value of $\sim 86$ nm at C$_{T}$ = 4.0 $\mu$M. For higher
concentrations, it abruptly decays to $\sim 43$ nm.

\begin{figure}
\includegraphics[width=8.5cm]{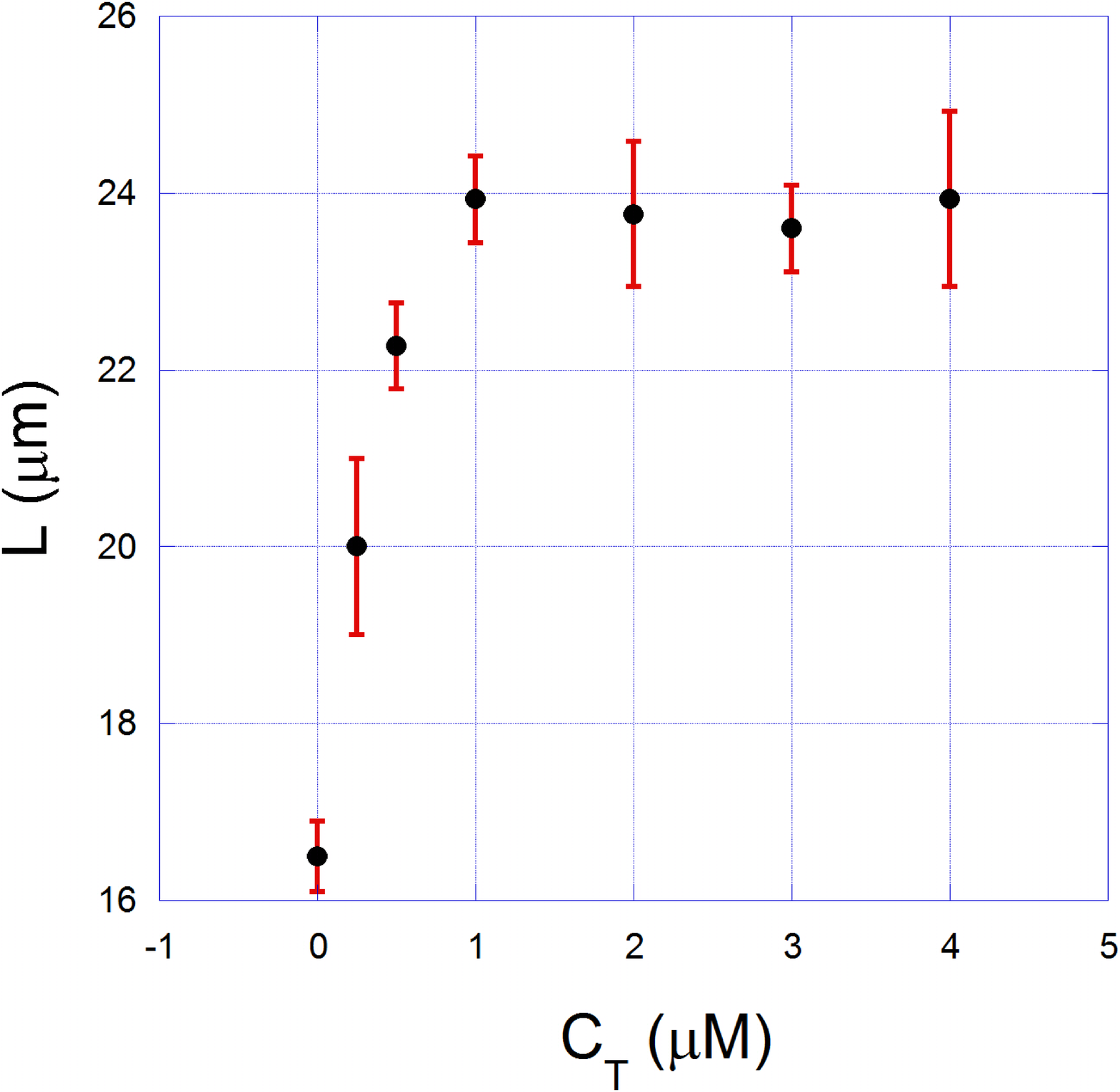}
\caption{Contour length $L$ of DNA-GelRed complexes as a function of
ligand total concentration in the sample $C_T$. Observe that $L$
monotonically increases from the bare DNA value ($\sim 16.5 \mu$m)
up to a saturation value of $\sim 24 \mu$m.}\label{LxCt}
\end{figure}

In Fig. \ref{LxCt} we show the behavior of the contour length $L$ of
DNA-GelRed complexes as a function of ligand total concentration in
the sample $C_T$. Observe that $L$ monotonically increases from the
bare DNA value ($\sim 16.5 \mu$m) up to a saturation value of $\sim
24 \mu$m.

To achieve these results, the experiments were performed as follows.
Firstly, we choose a particular bare DNA molecule and stretch it
five times, determining the mean values of the persistence and
contour lengths. Then, we change the ligand concentration in the
sample using a micropipette. After changing the concentration, we
wait $\sim 20$ minutes for the ligand to equilibrate with DNA. This
time scale is sufficient for GelRed to equilibrate with DNA for our
experimental conditions. This fact was verified by analyzing the
reversibility of the stretching curves and by performing some
experiments waiting longer time intervals, finding no significant
difference. We then perform again five stretching experiments,
obtaining the new values of the persistence and contour lengths for
the chosen GelRed concentration. This procedure is then repeated
sequentially for each ligand concentration. We have also repeated
the entire procedure, scanning all the concentrations, for other DNA
molecules using different samples. The error bars presented in Fig.
\ref{AxCt} and Fig. \ref{LxCt} for each concentration are the
standard errors obtained from this set of measurements.

The behavior of both persistence and contour lengths strongly
suggests that the dominant mechanism of interaction between DNA and
GelRed is the intercalative binding. In fact, the general behavior
of the persistence length $A$ shown in Fig. \ref{AxCt} was
previously verified by our group for the intercalators ethidium
bromide (EtBr), daunomycin, psoralen and diaminobenzidine under
nearly similar experimental conditions \cite{RochaJCP2},
\cite{RochaPB}, \cite{RochaAPLPso}, \cite{Reis}, by using optical
tweezers in the low-force regime ($\leq$ 2 pN). Some authors have
also reported the same qualitative behavior of the persistence
length for intercalators, finding that this parameter increases for
low ligand concentrations and decreases for higher concentrations
\cite{Tessmer}, \cite{Cassina}, \cite{Kaji}. In refs.
\cite{RochaPB}, \cite{RochaAPLPso} it was proposed that the abrupt
decrease of the persistence length shown in Fig. \ref{AxCt} for $C_T
>$ 4 $\mu$M is related to partial DNA denaturation, with the
formation of denaturing bubbles, probably due to the pulling force
\cite{RochaJCP2}, \cite{RochaPB}, \cite{RochaAPLPso}, \cite{Reis}.

In addition, it is well established that intercalators always
increase the DNA contour length when binding, by increasing the
axial distance between two adjacent base-pairs in the intercalative
site \cite{Sischka}, \cite{Fritzsche}, \cite{Chaires}. The other
common types of interactions between DNA and ligands, such as groove
binding, electrostatic interaction or covalent binding, do not
increase the DNA contour length. On the contrary, in some cases
these kinds of interaction can cause DNA compaction with a decrease
of the ``apparent contour length'' measured by force spectroscopy in
the low-force regime used in our experiments \cite{CrisafuliIB},
\cite{Silva}. Therefore, due to the strong increase of $L$ for the
DNA-GelRed complexes, we can conclude that if there exist other
binding mode, it is certainly much weaker than intercalation.

\begin{figure}
\includegraphics[width=8.5cm]{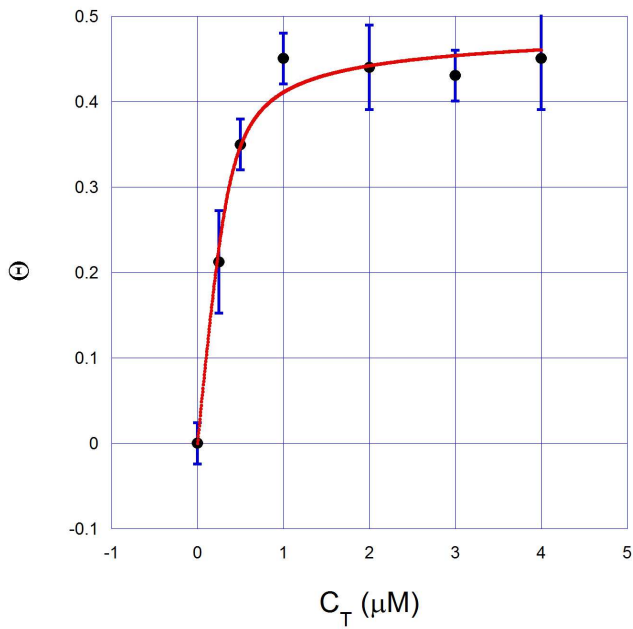}
\caption{Fractional increase of the contour length, $\Theta$ = ($L$
- $L_{0}$)/$L_{0}$ as a function of ligand concentration $C_T$.
\textit{Circles}: experimental data. \textit{Solid line}: a fitting
to the McGhee-von Hippel binding isotherm. The fitting returns the
physicochemical parameters $n$ = 3.8 $\pm$ 0.5, $K_i$ = (2.4 $\pm$
0.3) $\times$ 10$^7$ M$^{-1}$, and $\gamma$ = 1.9 $\pm$
0.1.}\label{bindingGelRed}
\end{figure}

The physicochemical properties of the interaction can be extracted
from the contour length data. Firstly, we use the data of Fig.
\ref{LxCt} to determine the fractional increase of the contour
length, $\Theta$ = ($L$ - $L_{0}$)/$L_{0}$, where $L_{0}$ is the
bare DNA contour length, as a function of ligand concentration
$C_T$. This data is plotted in Fig. \ref{bindingGelRed}
(\textit{circles}). For intercalators this fractional increase is
proportional to the bound ligand fraction $r$ = $C_b$/$C_{bp}$,
while $C_b$ is the bound ligand concentration and $C_{bp}$ is the
DNA base-pair concentration. In other words, $\Theta = \gamma r$,
where $\gamma$ is the ratio between the increase of the contour
length due to a single intercalating molecule ($\delta$) and the
mean distance between two consecutive base pairs in the bare DNA
($\Delta \sim$ 0.34 nm) \cite{RochaJCP2}, \cite{RochaPB},
\cite{Daune}. For typical monointercalating molecules one has
$\delta \sim$ 0.34 nm and consequently $\gamma \sim$ 1
\cite{Sischka}, \cite{Fritzsche}. The bound fraction $r$ (and
consequently the mechanical parameter $\Theta$) can be linked to the
physicochemical properties by using a binding isotherm. The McGhee -
von Hippel binding isotherm usually describes very well the physical
chemistry of the DNA interactions with intercalators, since it
computes in detail the neighbor exclusion effects which are usually
associated to this type of interaction \cite{McGhee},
\cite{RochaNEM}. This binding isotherm reads

\begin{equation}
\frac{r}{C_f} = K_i(1 - nr)\left[\frac{1 - nr}{1 - (n -
1)r}\right]^{n - 1} \label{nemfin},
\end{equation}
where $n$ is the exclusion number (a measure of the effective number
of base-pairs occupied by a single ligand molecule \cite{McGhee}),
$K_i$ is the intrinsinc equilibrium association constant and $C_f$ =
$C_T$ - $C_b$ is the free (not bound) ligand concentration in
solution.

In Fig. \ref{bindingGelRed} we show a fitting (\textit{solid line})
of this binding isotherm to the experimental data, by using the
computational method described in detail in ref. \cite{Cesconetto}.
Observe that the model describes well the experimental data. The
fitting returns the parameters $n$ = 3.8 $\pm$ 0.5, $K_i$ = (2.4
$\pm$ 0.3) $\times$ 10$^7$ M$^{-1}$, and $\gamma$ = 1.9 $\pm$ 0.1.
These results strongly suggest that the GelRed dye is a DNA
bis-intercalator. In fact, the exclusion parameter indicates that
each bound GelRed molecule effectively occupies 3.8 DNA base-pairs,
a value considerably higher than the results found for most
monointercalators, and approximately twice the result for EtBr
\cite{RochaPB}, \cite{Chaires}, \cite{Gaugain}. The equilibrium
constant is also higher than the result obtained for typical
monointercalators ($\sim$ 10$^5$ M$^{-1}$) \cite{RochaJCP2},
\cite{RochaPB}, \cite{Chaires}, \cite{Gaugain}, and within the range
found for most bis-intercalators (10$^7$ to 10$^9$ M$^{-1}$)
\cite{Gunther}, \cite{Berge}, \cite{Murade}, \cite{Maaloum},
\cite{Garbay}. In particular, it is two orders of magnitude higher
than the equilibrium constant for EtBr, a situation very similar to
what occurs for the bis-intercalator YOYO when compared to its
precursor YO, a monointercalator as EtBr \cite{Murade}. Finally, the
result $\gamma$ = 1.9 $\pm$ 0.1 is approximately twice the value
obtained for typical monointercalators, suggesting that each bound
GelRed molecule increases the DNA contour length by $\sim$ 0.65 nm,
a result also compatible to typical bis-intercalators
\cite{Gunther}, \cite{Maaloum}. Observe that the bis-intercalators
should increase approximately twice the DNA contour length per bound
molecule, since each ligand molecule contains two intercalating
portions.

These results together are a strong evidence that the GelRed dye is
in fact a bis-intercalator probably consisting of two EtBr
molecules, as claimed before \cite{WikiGelRed}. In fact, if one
supposes that GelRed is really a bis-intercalator formed by
crosslinking two EtBr molecules, it is straightforward to understand
the statement of the dye manufacturer which claims that GelRed
presents a higher sensitivity in electrophoresis experiments when
compared to EtBr \cite{GRSafety}: if one prepares two
electrophoresis assays staining one of them with EtBr and the other
with GelRed at the same molar concentrations, the GelRed assay will
present approximately twice DNA bound sites (at least for
concentrations far from saturation), which implies in more
fluorescence signal and consequently more contrast. In addition, the
fact that the absorption/emission spectra of the two compounds are
the same \cite{GRSafety} can be easily understood with the
discussion above.

\subsection{DLS experiments}

\begin{figure}
\centering
\includegraphics[width=10cm]{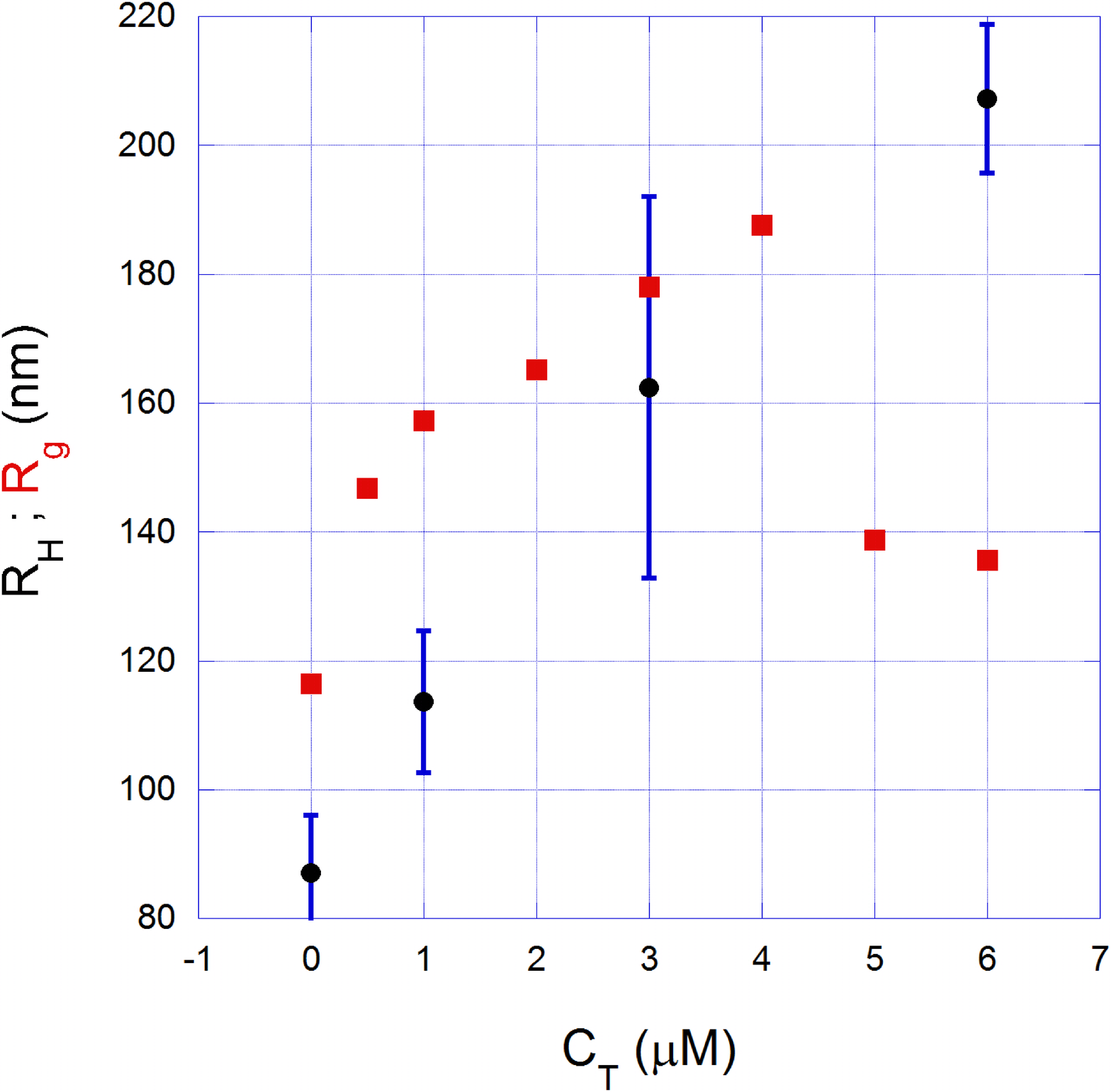}
\caption{\textit{Circles}: hydrodynamic radius $R_H$ of DNA-GelRed
complexes, obtained from the DLS experiments, as a function of
GelRed concentration $C_T$. \textit{Squares}: an estimative of the
radius of gyration $R_g$ for the DNA-GelRed complexes, obtained from
the data of the stretching experiments. The fact that $R_H$
increases monotonically with the ligand concentration indicates that
the abrupt transition of the persistence length do not occur in the
samples used for the DLS experiments.} \label{RhRgGelRed}
\end{figure}

In Fig. \ref{RhRgGelRed} (\textit{circles}) we show the behavior of
the hydrodynamic radius $R_H$ of DNA-GelRed complexes, obtained from
the DLS experiments, as a function of GelRed concentration. Observe
that $R_H$ increases monotonically with GelRed concentration, from
$\sim$ 87 nm measured for bare DNA, up to $\sim$ 207 nm obtained for
$C_T$ = 6 $\mu$M. Each experimental point is the mean value
calculated from a set of $\sim$ 100 measurements of 15 seconds long,
and the error bars are the standard deviations. In the same figure,
we represent an estimation of the radius of gyration $R_g$ for the
DNA-GelRed complexes (\textit{squares}). $R_g$ was calculated as a
function of the mechanical parameters obtained from the tweezers
experiments as $R_g = \sqrt{\frac{1}{3}AL\left(1 - \frac{3A}{L} +
...\right)}$ \cite{Daune}. We have used the values of the
persistence length $A$ shown in Fig. \ref{AxCt} and have assumed
that the contour length $L$ of the 3000 bp DNA will increase with
the ligand concentration on the same ratio of that shown in Fig.
\ref{bindingGelRed} for the $\lambda$-DNA. As pointed before in ref.
\cite{Reis}, since the 3000 bp DNA is sufficiently long (it contains
$\sim$ 20 persistence lengths), it is reasonable to expect that
base-pair sequence and other molecular details do not interfere on
large-scale mechanical properties such as the persistence and
contour lengths.

The radius of gyration $R_g$ increases with both $A$ and $L$. Thus,
as shown in Fig. \ref{RhRgGelRed}, $R_g$ decreases for the two
largest concentrations used, due to the abrupt decrease of the
persistence length shown in Fig. \ref{AxCt}. A similar decrease was
not verified for the hydrodynamic radius $R_H$, indicating that the
abrupt transition of the persistence length do not occur in the
samples used for the DLS experiments. The same result was previously
verified by our group for the intercalator diaminobenzidine
\cite{Reis}. As discussed in this previous work, such result was
expected since the abrupt transition of the persistence length is
probably related to the pulling force exerted in the stretching
experiments, which may locally denature the previously deformed
double-helix structure of the DNA-intercalator complex
\cite{RochaPB}, \cite{RochaAPLPso}. In this way, one can say that
the behavior of $R_H$ obtained in DLS experiments qualitatively
agrees with the results obtained from the optical tweezers
experiments. In addition, our DLS results also agree with the
results obtained by most authors who have measured the persistence
length of DNA-intercalator complexes with non-stretching techniques
(fluorescence microscopy, electron microscopy, viscosimetry, etc)
\cite{Yoshikawa92}, \cite{Quake}. These authors have found that
intercalators, aside increasing the DNA contour length, in general
also increase the DNA persistence length under these conditions,
thus increasing the effective size of the DNA-ligand complexes.
Nevertheless, it is important to mention that some works performed
with typical DNA-stretching techniques (optical or magnetic
tweezers) have found a monotonic decrease of the DNA persistence
length as a function of the intercalator concentration
\cite{Murade}, \cite{Sischka}, \cite{Lipfert}. In our opinion such
results are due to the force regime used to perform the
measurements, since as the forces used to stretch the DNA increase,
the probability of forming denaturation bubbles in the highly
distorted double-helix of the DNA-intercalator complexes increases
accordingly, thus decreasing the persistence length. Other factors
that can surely influence the results of such measurements are the
salt concentration used in the buffers, the model used to fit
force-extension data (which may include DNA stretch modulus if
forces are $\geq$ 10 pN), and the ratio of ligand concentration per
DNA base pair concentration \cite{RochaJCP2}.

\section{Conclusion}

By using two very different experimental techniques (single molecule
stretching and dynamic light scattering) we have characterized the
interaction of the DNA molecule with the fluorescent stain GelRed,
determining the changes of the mechanical properties of DNA-GelRed
complexes as a function of ligand concentration and extracting the
physical chemistry of the interaction from these data. It was found
that GelRed binds strongly to DNA ($K_i \sim$ 10$^7$ M$^{-1}$). In
addition, we have estimated that each bound GelRed molecule
effectively occupies $\sim$ 3.8 DNA base-pairs and increases the
contour length by $\sim$ 0.65 nm. These numbers are compatible to
the results expected for bis-intercalating molecules, which has
allowed us to determine the main binding mechanism of the GelRed dye
and to understand the higher sensitivity presented by this compound
when compared to ethidium bromide in electrophoresis experiments.

\begin{acknowledgements}
This work was supported by the Brazilian agencies: Funda\c{c}\~ao de
Amparo \`a Pesquisa do Estado de Minas Gerais (FAPEMIG) and Conselho
Nacional de Desenvolvimento Cient\'ifico e Tecnol\'ogico (CNPq). The
DLS experiments were performed in ``Laborat\'orio de Microflu\'idica
e Fluidos Complexos (LMFFC)'' of Universidade Federal de Vi\c{c}osa.
\end{acknowledgements}

\bibliographystyle{spmpsci}      

\bibliography{gelred_bibtex}

\begin{thebibliography}{10}
\providecommand{\url}[1]{{#1}}
\providecommand{\urlprefix}{URL }
\expandafter\ifx\csname urlstyle\endcsname\relax
  \providecommand{\doi}[1]{DOI~\discretionary{}{}{}#1}\else
  \providecommand{\doi}{DOI~\discretionary{}{}{}\begingroup
  \urlstyle{rm}\Url}\fi

\bibitem{WikiGelRed}
Gelred.
\newblock http://en.wikipedia.org/wiki/GelRed

\bibitem{GRSafety}
Safety report of gelred and gelgreen.
\newblock Nucleic Acid Detection Technologies - http://www.biotium.com

\bibitem{Amitani}
Amitani, I., Liu, B., Dombrowski, C.C., Baskin, R.J., Kowalczykowski, S.C.:
  Watching individual proteins acting on single molecules of dna.
\newblock Methods Enzymol. \textbf{472}, 261--291 (2010)

\bibitem{Berge}
Berge, T., Jenkins, N.S., Hopkirk, R.B., Waring, M.J., Edwardson, J.M.,
  Henderson, R.M.: Structural perturbations in dna caused by bis-intercalation
  of ditercalinium visualised by atomic force microscopy.
\newblock Nucl. Acids Res. \textbf{30}(13), 2980--2986 (2002)

\bibitem{Cassina}
Cassina, V., Seruggia, D., Beretta, G.L., Salerno, D., Brogioli, D., Manzini,
  S., Zunino, F., Mantegazza, F.: Atomic force microscopy study of dna
  conformation in the presence of drugs.
\newblock Eur. Biophys. J. \textbf{40}(1), 59--68 (2010)

\bibitem{Cesconetto}
Cesconetto, E.C., Junior, F.S.A., Crisafuli, F.A.P., Mesquita, O.N., Ramos,
  E.B., Rocha, M.S.: Dna interaction with actinomycin d: Mechanical
  measurements reveal the details of the binding data.
\newblock Phys. Chem. Chem. Phys. \textbf{15}(26), 11,070--11,077 (2013)

\bibitem{Chaires}
Chaires, J.B., Dattagupta, N., Crothers, D.M.: Studies on interaction of
  anthracycline antibiotics and deoxyribonucleic-acid - equilibrium
  binding-studies on interaction of daunomycin with deoxyribonucleic-acid.
\newblock Biochemistry \textbf{21}(17), 3933--3940 (1982)

\bibitem{CrisafuliIB}
Crisafuli, F.A.P., Cesconetto, E.C., Ramos, E.B., Rocha, M.S.: Dna-cisplatin
  interaction studied with single molecule stretching experiments.
\newblock Integr. Biol. \textbf{2012}(4), 568--574 (2012)

\bibitem{Daune}
Daune, M.: Molecular Biophysics, first edn.
\newblock Oxford University Press, Oxford (1999)

\bibitem{Fritzsche}
Fritzsche, H., Triebel, H., Chaires, J.B., Dattagupta, N., Crothers, D.M.:
  Interaction of anthracycline antibiotics with bio-polymers .6. studies on
  interaction of anthracycline antibiotics and deoxyribonucleic-acid - geometry
  of intercalation of iremycin and daunomycin.
\newblock Biochemistry \textbf{21}(17), 3940--3946 (1982)

\bibitem{Garbay}
Garbay-Jaureguiberry, C., Laug\^{a}a, P., Delepierre, M., Laalami, S., Muzard,
  G., Pecq, J.B.L., Roques, B.P.: Dna bis-intercalators as new anti-tumour
  agents: Modulation of the anti-tumour activity by the linking chain rigidity
  in the ditercalinium series.
\newblock Anticancer Drug Des. \textbf{1}(4), 323--335 (1987)

\bibitem{Gaugain}
Gaugain, B., Barbet, J., Capelle, N., Roques, B.P., Pecq, J.L.: Dna
  bifunctional intercalators .2. fluorescence properties and dna binding
  interaction of an ethidium homodimer and an acridine ethidium heterodimer.
\newblock Biochemistry \textbf{17}(24), 5078--5088 (1978)

\bibitem{Gunther}
G\"{u}nther, K., Mertig, M., Seidel, R.: Mechanical and structural properties
  of yoyo-1 complexed dna.
\newblock Nucl. Acids Res. \textbf{38}(19), 6526--6532 (2010)

\bibitem{Huang}
Huang, Q., Baum, L., Fu, W.L.: Simple and practical staining of dna with gelred
  in agarose gel electrophoresis.
\newblock Clin. Lab. \textbf{56}, 149--152 (2010)

\bibitem{Hur}
Hur, J.S., Shaqfeh, E.S.G.: Dynamics of dilute and semidilute dna solutions in
  the start-up of shear flow.
\newblock J. Rheol. \textbf{45}(2), 421--450 (2001)

\bibitem{Kaji}
Kaji, N., Ueda, M., Baba, Y.: Direct measurement of conformational changes on
  dna molecule intercalating with a fluorescence dye in an electrophoretic
  buffer solution by means of atomic force microscopy.
\newblock Electrophoresis \textbf{22}(16), 3357--3364 (2001)

\bibitem{Lipfert}
Lipfert, J., Klijnhout, S., Dekker, N.H.: Torsional sensing of small-molecule
  binding using magnetic tweezers.
\newblock Nucl. Acids Res. \textbf{38}(20), 7122--7132 (2010)

\bibitem{Maaloum}
Maaloum, M., Mullera, P., Harlepp, S.: Dna-intercalator interactions:
  Structural and physical analysis using atomic force microscopy in solution.
\newblock Soft Matter \textbf{9}, 11,233 (2013)

\bibitem{Marko}
Marko, J.F., Siggia, E.D.: Stretching dna.
\newblock Macromolecules \textbf{28}(26), 8759--8770 (1995)

\bibitem{McGhee}
McGhee, J.D., von Hippel, P.H.: Theoretical aspects of dna-protein interactions
  - cooperative and non-cooperative binding of large ligands to a
  one-dimensional homogeneous lattice.
\newblock J. Mol. Biol. \textbf{86}(2), 469--489 (1974)

\bibitem{Murade}
Murade, C.U., Subramaniam, V., Otto, C., Bennink, M.L.: Interaction of oxazole
  yellow dyes with dna studied with hybrid optical tweezers and fluorescence
  microscopy.
\newblock Biophys. J. \textbf{97}, 835--843 (2009)

\bibitem{Quake}
Quake, S.R., Babcock, H., Chu, S.: The dynamics of partially extended single
  molecules of dna.
\newblock Nature \textbf{388}(10), 151--154 (1997)

\bibitem{Reis}
Reis, L.A., Ramos, E.B., Rocha, M.S.: Dna interaction with diaminobenzidine
  studied with optical tweezers and dynamic light scattering.
\newblock J. Phys. Chem. B \textbf{117}(46), 14,345--14,350 (2013)

\bibitem{RochaPB}
Rocha, M.S.: Modeling the entropic structural transition of dna complexes
  formed with intercalating drugs.
\newblock Phys. Biol. \textbf{6}, 036,013 (2009)

\bibitem{RochaNEM}
Rocha, M.S.: Revisiting the neighbor exclusion model and its applications.
\newblock Biopolymers \textbf{93}, 1--7 (2010)

\bibitem{RochaJCP2}
Rocha, M.S., Ferreira, M.C., Mesquita, O.N.: Transition on the entropic
  elasticity of dna induced by intercalating molecules.
\newblock J. Chem. Phys. \textbf{127}(10), Art. 105,108 (2007)

\bibitem{RochaAPLPso}
Rocha, M.S., L\'ucio, A.D., Alexandre, S.S., Nunes, R.W., Mesquita, O.N.:
  Dna-psoralen: Single-molecule experiments and first principles calculations.
\newblock Appl. Phys. Lett. \textbf{95}, 253,703 (2009)

\bibitem{Silva}
Silva, E.F., Ramos, E.B., Rocha, M.S.: Dna interaction with hoechst 33258:
  stretching experiments decouple the different binding modes.
\newblock J. Phys. Chem. B \textbf{117}(24), 7292--6 (2013)

\bibitem{Siman}
Siman, L., Carrasco, I.S.S., da~Silva, J.K.L., Oliveira, M.C., Rocha, M.S.,
  Mesquita, O.N.: Quantitative assessment of the interplay between
  dna-elasticity and cooperative binding of ligands.
\newblock Phys. Rev. Lett. \textbf{109}(24), 248,103 (2012)

\bibitem{Sischka}
Sischka, A., Toensing, K., Eckel, R., Wilking, S.D., Sewald, N., Rios, R.,
  Anselmetti, D.: Molecular mechanisms and kinetics between dna and dna binding
  ligands.
\newblock Biophys. J. \textbf{88}(1), 404--411 (2005)

\bibitem{Tessmer}
Tessmer, I., Baumann, C.G., Skinner, G.M., Molloy, J.E., Hoggett, J.G.,
  Tendler, S.J.B., Allen, S.: Mode of drug binding to dna determined by optical
  tweezers force spectroscopy.
\newblock J. Mod. Opt. \textbf{50}(10), 1627--1636 (2003)

\bibitem{Yoshikawa92}
Yoshikawa, K., Matsuzawa, Y., Minagawa, K., Doi, M., Matsumoto, M.: Opposite
  effect between intercalator and minor groove binding drug on the higher order
  structure of dna as is visualized by fluorescence microscopy.
\newblock Bioch. Biphys. Res. Commun. \textbf{188}(3), 1274--1279 (1992)

\end{thebibliography}

\end{document}